\documentclass[aps,prb,twocolumn,superscriptaddress,showpacs]{revtex4}

\usepackage{graphicx}
\usepackage{natbib}

\newcommand{\gtappr}{{{\lower4pt\hbox{$>$} } \atop \widetilde{ \ \ \ }}}
\newcommand{\ltappr}{{{\lower4pt\hbox{$<$} } \atop \widetilde{ \ \ \ }}}

\newcommand{\beq}{\begin{equation}}
\newcommand{\eeq}{\end{equation}}

\newcommand{\ie}{\textit{i.e. }}
\newcommand{\etal}{\emph{et al.}}
                            
\newcommand{\ybal}{$\beta$-YbAlB$_4\,$}
\newcommand{\aybal}{$\alpha$-YbAlB$_4\,$}

\def\gsim{\buildrel {\textstyle >}\over {_\sim}}
\def\lsim{\buildrel {\textstyle <}\over {_\sim}}

\newsavebox{\fmbox}

\begin{document}


\title{Anisotropic heavy Fermi-liquid formation in the valence fluctuating \aybal}


\author{Yosuke Matsumoto}
\email[]{matsumoto@issp.u-toyko.ac.jp}
\affiliation{Institute for Solid State Physics, University of Tokyo, Kashiwa, Chiba 277-8581, Japan}

\author{K. Kuga}
\affiliation{Institute for Solid State Physics, University of Tokyo, Kashiwa, Chiba 277-8581, Japan}

\author{T. Tomita}
\thanks{Present address: 
Department of Physics, College of Humanities and Sciences, Nihon University, Sakurajosui, Setagaya-ku, Tokyo 156-8550, Japan}
\affiliation{Institute for Solid State Physics, University of Tokyo, Kashiwa, Chiba 277-8581, Japan}

\author{Y. Karaki}
\affiliation{Institute for Solid State Physics, University of Tokyo, Kashiwa, Chiba 277-8581, Japan}
\affiliation{Faculty of Education, University of the Ryukyus, Nishihara, Okinawa 903-0213, Japan}

\author{S. Nakatsuji}
\email[]{satoru@issp.u-tokyo.ac.jp}
\affiliation{Institute for Solid State Physics, University of Tokyo, Kashiwa, Chiba 277-8581, Japan}


\date{\today}

\begin{abstract}
\aybal is the locally isostructural polymorph 
of \ybal, the first example of an Yb-based heavy fermion superconductor  
which exhibits pronounced non-Fermi-liquid behavior above $T_{\rm c}$. 
Interestingly, both \aybal and \ybal have strongly intermediate valence. 
Our single crystal study of the specific heat, magnetization and resistivity has confirmed 
the Fermi liquid ground state of \aybal ~in contrast with the quantum criticality observed in \ybal. 
Both systems exhibit Kondo lattice behavior with the characteristic temperature scale $T^* \sim$ 8 K 
in addition to a valence fluctuation scale $\sim 200$ K. 
Below $T^*$, \aybal forms a heavy Fermi liquid state with an electronic specific heat coefficient $\gamma\sim$ 130 mJ/mol K$^2$ and 
a large Wilson ratio more than 7, which indicates ferromagnetic correlation between Yb moments.
A large anisotropy in the resistivity suggests that 
the hybridization between 4$f$ and conduction electrons is much stronger in the $ab$-plane than 
along the $c$-axis. 
The strongly anisotropic hybridization as well as the large Wilson ratio 
is the key to understand the unusual Kondo lattice behavior and heavy fermion formation in mixed valent compounds.
\end{abstract}


\maketitle


\section{Introduction}
$4f$ based heavy fermion (HF) systems have attracted much attention 
with interesting phenomena such as unconventional superconductivity and non-Fermi-liquid (NFL) behavior 
found in the vicinity of quantum critical points~\cite{Mathur98, Stewart01, Yuan03, Lohneysen07, Monthoux07, gegenwart08}.
Our recent studies have found 
the first Yb (4$f^{13}$) based HF superconductivity 
with the transition temperature $T_{\rm c}$ = 80 mK 
in the new compound $\beta$-YbAlB$_4$ \cite{nakatsuji08,KugaPRL}. 
Pronounced NFL behavior above $T_c$ and its magnetic field dependence indicate 
that the system is a rare example of a pure metal that displays quantum criticality 
at ambient pressure and close to zero magnetic field \cite{nakatsuji08}. 
Furthermore, the $T/B$ scaling found in our recent high-precision magnetization measurements 
clarifies its unconventional zero-field quantum criticality without tuning\cite{matsumoto-ZFQCP}, 
which can not be explained by the standard theory 
based on spin-density-wave fluctuations~\cite{Hertz76, Moriya85, Millis93}. 
In contrast to the canonical quantum critical materials, hard X-ray photoemission spectroscopy (HXPES) measurements have revealed strongly 
intermediate valence of Yb$^{+2.75}$, providing the only example of quantum criticality in a mixed valent system~\cite{ybal-valency}. 
Whether the valence fluctuation is relevant for the mechanism of the quantum criticality and superconductivity is an interesting open question.

In this paper, we present the results of the specific heat, magnetization and resistivity measurements of \aybal, 
the locally isostructural polymorph of \ybal with different arrangement of 
distorted hexagons made of Yb atoms (space group: $Pbam$(\aybal ), $Cmmm$(\ybal))~\cite{fisk81, Macaluso07}. 
According to the HX-PES measurement~\cite{ybal-valency}, \aybal also has an intermediate valence of Yb$^{+2.73}$. 
The results indicate a Fermi liquid (FL) ground state for \aybal in contrast to the unconventional quantum criticality observed in \ybal.  
Interestingly, both systems exhibit Kondo lattice behavior with 
a small renormalized temperature scale of $T^*\sim 8$ K 
although both of them have a large valence fluctuation scale of $\sim 200$ K. 
Below $T^*$, \aybal forms a heavy Fermi liquid state with an electronic specific heat coefficient $\gamma\sim$ 130 mJ/mol K$^2$ and 
a large Wilson ratio more than 7, which indicates ferromagnetic correlation between Yb moments.
Kadowaki-Woods ratio is found similar to those found in the normal Kondo lattice systems and 
considerably larger than mixed valent systems. 
Furthermore, a large anisotropy observed in the resistivities of \aybal suggests strongly anisotropic hybridization between 4$f$ and conduction electrons.
This strong anisotropy in the hybridization is the key to understand the mechanism of the heavy fermion formation as well as 
the Kondo lattice behavior found in the intermediate valence system. 
Partial information has already been discussed in Ref.~\cite{matsumoto-SCES2010
}.  

\section{Experimental}
High purity single crystals of \aybal were grown by a flux method \cite{Macaluso07}. 
Energy dispersive X-ray (EDX) and induction coupled plasma (ICP) analyses found no impurity phases, no inhomogeneities and a ratio Yb:Al of 1:1. 
Surface impurities were carefully removed with dilute nitric acid before measurements. 
We succeeded in growing pure crystals with residual resistivity ratio (RRR) up to 110. 
The magnetization $M$ at $T > 2$ K was measured by a commercial SQUID magnetometer 
using pure single crystals (RRR $\sim$ 50) of 2.4 mg. 
The magnetization data at $T < 4$ K and 
$B < 0.05$ T were obtained by using a high precision SQUID magnetometer installed in a $^3$He-$^4$He dilution refrigerator \cite{matsumoto-ZFQCP}. 
The specific heat $C$ of pure single crystals ( 1.1 mg, RRR $\sim$ 50) 
was measured at temperature range $0.4 < T < 200$ K by a relaxation method 
using a physical property measurement system. 
Four-terminal resistivity
measurements were made by using a DC method (300 K $\gsim T \gsim$ 0.5 K) and an AC method (1.4 K $\gsim T \gsim$ 35 mK).

\section{Results and Discussion}
\begin{figure}[tb]
\begin{center}
\includegraphics[width = 7.6 cm,clip]{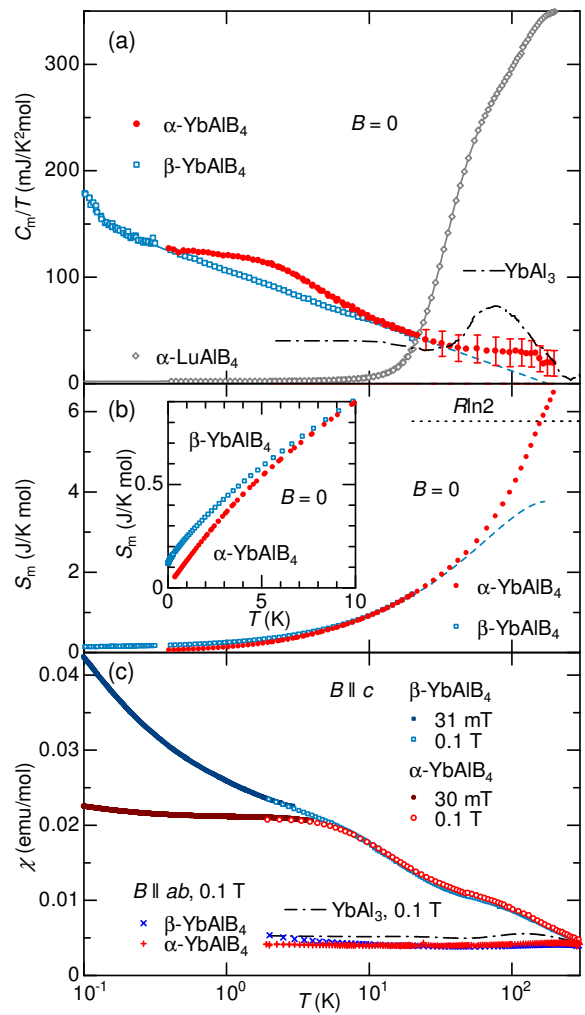}
\end{center}
\caption{(a) Magnetic part ($f$-electron contribution) of the specific heat $C_{\rm m}$ 
plotted as $C_{\rm m}/T$ versus $T$ for both $\beta$- and $\alpha$-YbAlB$_4$ under zero field. 
$C_{\rm m}/T$ for the $\beta$ phase shows a $\ln T$ dependence 
for 0.2 K $< T <$ 20 K~\cite{matsumoto-ZFQCP}. 
The broken line is the fit of the results to 
${C_{\rm m}}/{T} = {S_0}/{T_0}\ln\left({T_0}/{T} \right)$ (see text). 
$C/T$ of $\alpha$-LuAlB$_4$ and $C_{\rm m}/T$ of the intermediate valent cubic system YbAl$_3$~\cite{YbAl3} are also shown. 
(b) The magnetic part of the entropy $S_{\rm m}$, which was obtained by integrating $C_{\rm m}/T$. 
In \aybal, a constant value of 127 mJ/K$^2$mol is assumed below 0.4 K.  
Above 10 K, $C_{\rm m}/T$ in \aybal merges to the ln$T$ behavior of \ybal. 
Thus the data for \ybal was shifted to take the same value as the one of \aybal at 20 K. 
The broken line is obtained from the $\ln T$ fitting used in (a). 
Inset shows the low $T$ part.  
(c) Temperature dependence of the d.c. susceptibility $\chi = M/B$ 
measured in a field along the $ab$ plane and $c$-axis for both $\beta$- and $\alpha$-YbAlB$_4$. 
$\chi$ of the intermediate valent cubic system YbAl$_3$~\cite{YbAl3} is also shown. }
\label{f1}
\end{figure}

First, we present the magnetic part of the specific heat $C_{\rm m}$ divided by temperature in Fig. \ref{f1} (a). 
$C_{\rm m}$ was obtained by subtracting the specific heat of $\alpha$-LuAlB$_4$ shown in the same figure. 
Here, $\alpha$-LuAlB$_4$ is the non-magnetic isostructural counterpart of \aybal. 
The Debye temperature of $\alpha$-LuAlB$_4$ is estimated to be 380 K from the $T^3$ dependence of $C$ below 10 K. 
In both $\alpha$- and $\beta$-YbAlB$_4$, $C_{\rm m}/T$ is strongly enhanced 
to be $\gsim$ 130 mJ/molK$^2$ in the low $T$ limit,
which is large compared to ordinary valence fluctuating materials, 
such as CeSn$_3$\cite{CeSn3-gamma} and YbAl$_3$ \cite{YbAl3}(see Fig. \ref{f1} (a)) and 
is two orders magnitude larger than the band calculation estimates ($\sim 6$~mJ/molK$^2$)~\cite{Andriy09, Eoin_PRL09}. 
While clear ln$T$ divergent behavior is observed in \ybal in the temperature range of 0.2 K $< T <$ 20 K, 
$C_{\rm m}/T$ in \aybal 
nearly saturates at $T<$ 1 K, indicating a Fermi liquid ground state. 
On the other hand, at higher temperatures above 10 K, $C_{\rm m}/T$ in \aybal merges to the ln$T$ behavior of \ybal. 
Fitting the ln$T$ behavior of \ybal to ${C_{\rm m}}/{T} = {S_0}/{T_0}\ln\left({T_0}/{T} \right)$ 
yields $T_0 = 180 \pm 10$ K and $S_0 = 3.7 \pm 0.1$ J/molK for \ybal~\cite{matsumoto-ZFQCP}. 
Here, $T_0$ provides a characteristic hybridization scale for the system 
and is close to the coherence temperature of 250 K set by the resistivity peak~\cite{nakatsuji08}. 
Another rough estimate of $T_0$ can be made using the temperature where 
the magnetic part of the entropy $S_{\rm m}$ reaches $R\ln 2$ (the entropy of a ground state doublet). 
In this way, $T_0$ for \aybal can be estimated to be $T_0 \sim 160 \pm 20$ K, 
as shown in Fig. \ref{f1} (b). 
In order to obtain $S_{\rm m}$, we assume a constant value of $C_{\rm m}/T$ (127 mJ/molK$^2$) below the lowest temperature of the measurements 0.4 K. 
These large values of $T_0$ are consistent with the intermediate valence of these systems because  
mixed-valent compounds are typically characterized by a much higher value of 
$T_0$ than Kondo lattice systems~\cite{lohneysen96,gegenwart08,Custers03}. 
A proposed crystalline electric field (CEF) level scheme, which reproduces the magnetic susceptibility, 
suggests a CEF level splitting of $\Delta$ = 80 K~\cite{Andriy09}. 
However, a Schottky peak of this level splitting which would appear at $\sim 25$ K with a height of 130 mJ/$K^2$mol is not seen here.
This is probably because the CEF levels are smeared out by the valence fluctuations.

\begin{figure}[tb]
\begin{center}
\includegraphics[width = 7.3 cm, clip]{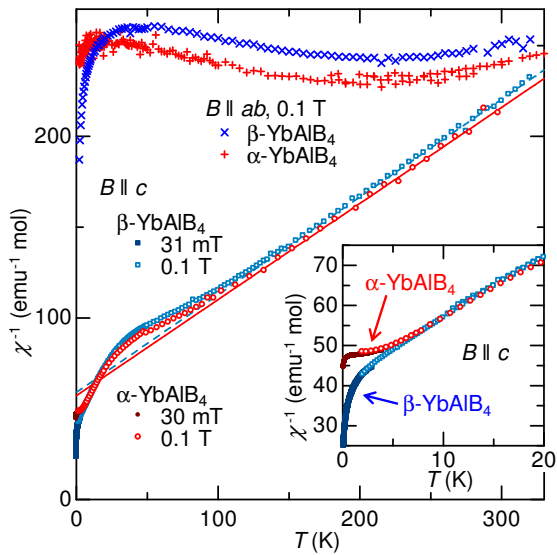}
\end{center}
\caption{Temperature dependence of the inverse susceptibility $\chi ^{-1} = B/M$
under the field along the the $ab$-plane and $c$-axis. 
Solid and broken lines are Curie-Weiss fits above 150 K for 
$\alpha$- (open circles) and \ybal (open squares), respectively. 
Inset shows the low temperature part of $\chi ^{-1}$ under the field along the the $c$-axis.}
\label{f2}
\end{figure}

The temperature dependence of the d.c. magnetic susceptibility $\chi = M/B$ are shown in Fig. \ref{f1} (c). 
Both systems exhibit strong Ising anisotropy with the strongly $T$ dependent $c$-axis $\chi$ and  
almost $T$ independent $\chi$ along the $ab$-plane~\cite{Macaluso07}. 
Broad peaks found around 200 K in $\chi _{ab}$ for both systems (Fig. \ref{f2})
are close to the $T_0$ scale obtained from $C_{\rm m}$ and the coherence temperature of the resistivity which we will discuss later. 
The $c$-axis component for both systems shows almost the same temperature dependence down to $T \sim 8$ K.
~Below $T \lsim 8$ K, on the other hand, these two systems show contrasting behavior: while 
\ybal continues to diverge due to the quantum criticality \cite{matsumoto-ZFQCP}, 
\aybal shows saturating behavior, indicating the Fermi liquid formation. 
The Curie-Weiss behavior, $\chi _c = C/(T+\Theta_{\rm W})$, is observed at $T > 150$ K 
with $\Theta_{\rm W} = 110 \pm 2$, $108\pm 5$ K 
for $\alpha$ and $\beta$ phases, respectively (Fig.~\ref{f2}). 
Ising moments $I_{\rm z} = 2.22$, 2.24 $\mu _{\rm B}$ for $\alpha$ and $\beta$ phases 
are deduced from the Curie constant $C = N_{\rm A}I_{\rm z}^2/k_{\rm B}$ where $N_{\rm A}$ and $k_{\rm B}$ are Avogadro and Boltzmann constants, respectively.
Furthermore, at $T< 20$ K, another Curie-Weiss behavior 
is observed (Fig.~\ref{f2} inset). 
If we fit the data to the Curie-Weiss law at $6 \lsim T \lsim 15$ K, 
$\Theta_{\rm W}= 29$, 25 K and $I_{\rm z} =$ 1.4, 1.3 $\mu _{\rm B}$ are obtained for the $\alpha$ and $\beta$ phases, respectively. 

These observations suggest existence of local moments far below $T_0 \sim 200$~K possibly down to $ \sim 8$~K. 
This Kondo lattice behavior with a low temperature scale $T^*\sim 8$~K is 
striking compared with ordinary valence fluctuating materials where Pauli paramagnetism is normally expected, 
such as CeSn$_3$~\cite{tsuchida_CeSn3} and YbAl$_3$ \cite{YbAl3}(see Fig. \ref{f1} (c)). 
One of the possible origins of this behavior may lie in Kondo resonance narrowing~\cite{kondo-narrowing} 
due to the presence of ferromagnetic (FM) interactions between Yb $4f$-electron spins where 
FM interactions cause a large downward renormalization of the Kondo temperature from $T_0 \sim 200$~K to $T^*\sim 8$~K~\cite{matsumoto-ZFQCP}. 
Indeed, the Wilson ratio $R_{\rm W} = (\pi ^2 k_B^2/\mu _0I_{\rm z}^2)(\chi / \gamma )\sim 7$ 
is obtained for both $\alpha$ and $\beta$ phases 
by using $\chi_{\rm c}$ at $B = 0.1$ T, $T = 0.4$ K, $\gamma = C_{\rm m}/T$ at $B = 0$, $T = 0.4$ K 
and  $I_z$ obtained from the high temperature Curie-Weiss fit.   
The $R_{\rm W}$ values are considerably large compared with the normal value 2 expected for Kondo lattice systems. 
If we use $I_{\rm z}$ obtained from the low temperature Curie-Weiss fit, the Wilson ratio becomes 
$R_{\rm W}\sim 25$ for both systems, 
these significantly high values can be regarded as a consequence of the FM correlations. 

Alternatively, the large $R_{\rm W}$ values might also be explained by the  
possible proximity to a valence quantum criticality as it have been recently pointed out by Watanabe and Miyake~\cite{watanabe-PRL105}.
In this case, the low temperature scale  $T^*\sim 8$~K might arise from  
the characteristic energy scale for the valence fluctuations and not from the Kondo resonance narrowing. 
So far, we do not have experimental evidence to uniquely specify the mechanism among the possible scenarios. 
Further studies are required to solve this issue.

\begin{figure}[tb]
\begin{center}
\includegraphics[width = 8 cm, clip]{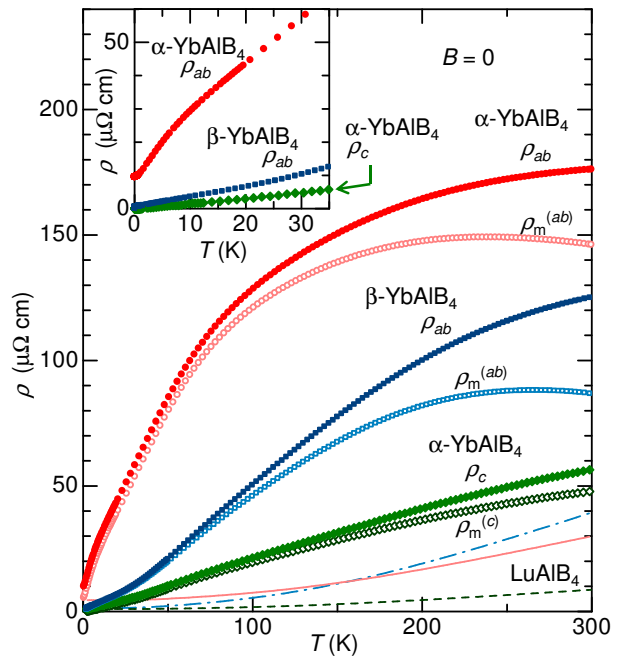}
\end{center}
\caption{Temperature dependence of the in-plane and $c$-axis resistivity $\rho _{ab}$ and $\rho _{c}$ of \aybal and 
$\rho _{ab}$ of \ybal. 
The magnetic part of the resistivity $\rho _{{\rm m}}$ is obtained by subtracting non-magnetic 
contribution estimated by $\rho _{ab}$ of $\alpha$-, $\beta$-LuAlB$_4$(solid and 
dash dotted lines, respectively) or $\rho _{c}$ of $\alpha$-LuAlB$_4$(dashed line). 
Inset shows low $T$ part of $\rho _{ab}$ and $\rho _{c}$. 
}
\label{f3}
\end{figure}

The temperature dependence of the in-plane resistivity with current along [-110] direction, 
which we denote $\rho _{ab}$, 
and $c$-axis resistivity $\rho _{c}$ are shown in Fig. \ref{f3}. 
We have also measured  the $a$-axis resistivity $\rho _{a}$ and have found no significant difference from $\rho _{ab}$. 
This is consistent with the  the recent band calculation~\cite{Tompsett10}, which predicts a nearly isotropic transport within the plane.
Further investigation of the in-plane anisotropy including the $b$-axis resistivity $\rho _{b}$ is now underway. 
Note that $\rho _{c}$ in \ybal is not available so far due to the tiny thickness of $\sim$ 10 $\mu$m along the $c$-axis of single crystals. 
The magnetic part of the resistivity $\rho _{{\rm m}}$ is obtained by subtracting 
the corresponding component of $\rho$ of the non-magnetic analog  
$\alpha$- or $\beta$-LuAlB$_4$. 
The in-plane magnetic component, $\rho _{{\rm m}}^{(ab)}$, exhibits broad peaks at $T\sim$ 250 K in 
both \aybal and \ybal, which are close to the peak temperatures of $\chi _{ab}(T)$ and $T_0$ obtained from $C_{\rm m}$.  
Therefore, these may be considered as the coherence peak providing the characteristic hybridization temperature scale. 
On the other hand, $\rho _{{\rm m}}^{(c)}$ in \aybal decreases monotonously on cooling below 300 K.

\begin{figure}[tb]
\begin{center}
\includegraphics[width = 8 cm, clip]{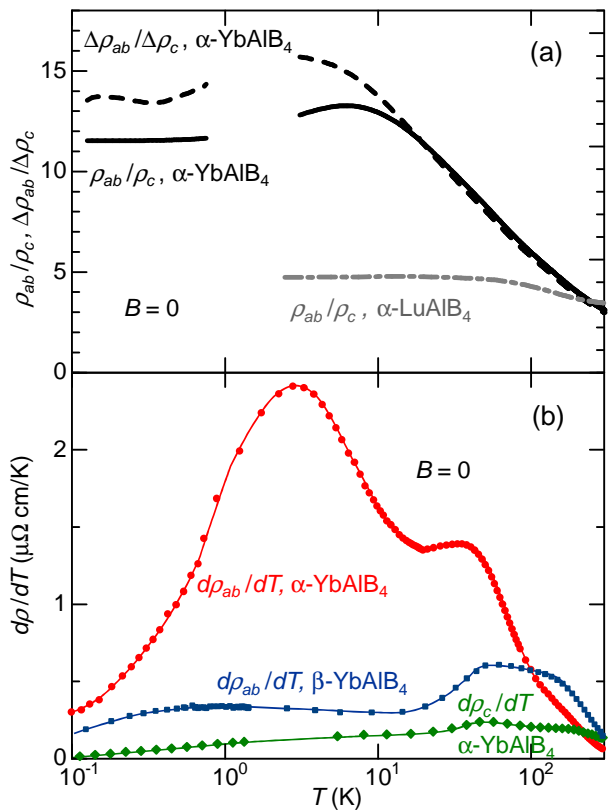}
\end{center}
\caption{(a) Temperature dependence of the ratio $\rho _{ab}$/$\rho _{c}$ and $\Delta\rho _{ab}$/$\Delta\rho _{c}$. 
Here $\Delta\rho$ is defined by $\Delta\rho\equiv\rho - \rho _{0}$ (see text).
(b) Temperature derivative of the resistivity $d\rho/dT$. 
}
\label{f4}
\end{figure}

Interestingly, $\rho _{c}$ is much smaller than $\rho _{ab}$ in \aybal 
~\ie the conductivity of the system exhibits quasi-1D anisotropy. 
The ratio, 
$\rho _{ab}$/$\rho _{c}$, increases at low temperatures making a peak at $T\sim$ 6 K (Fig. \ref{f4} (a), solid black line). 
At the peak, $\rho _{ab}$/$\rho _{c}$ reaches $\sim$ 13 and approaches to a constant value of $\sim$ 11 in the lowest temperatures.  
This low temperature anisotropy is one order of magnitude larger than typical anisotropic heavy fermion systems such as 
CeCoIn$_5$~\cite{Malinowski05}, CeCu$_2$Si$_2$~\cite{Onuki_JPSJ53}, CeNiIn~\cite{Fujii92}, 
YbAgGe~\cite{Niklowitz06} 
where the ratio is almost $T$ independent and $\lsim$ 3 below 300 K. 
On the other hand, $\rho _{ab}$/$\rho _{c}$ in $\alpha$-LuAlB$_4$ is nearly temperature independent 
with a slight increase from 3.5 at 300 K to 4.8 at lowest temperatures (Fig. \ref{f4} (a), dotted broken gray line). 
This temperature independent anisotropy in $\alpha$-LuAlB$_4$ should come from the anisotropy of the Fermi surface. 
Interestingly, at $T\sim$ 300 K, $\rho _{ab}$/$\rho _{c}$ in \aybal approaches to a similar value to the 
one in $\alpha$-LuAlB$_4$ although $f$ electron contribution is dominant in \aybal. 
This suggests that the topology of the Fermi surfaces of these systems are 
similar to each other at high $T>T_0$. 

The peak found in $\rho _{ab}$/$\rho _{c}$ arises mainly from a rapid decrease in 
$\rho _{ab}$ below $T\sim $ 10 K (Fig. \ref{f3} inset). 
This can be clearly seen in 
the temperature derivative of $\rho$, $d\rho/dT$ shown in Fig. \ref{f4} (b). 
While $d\rho _{c}/dT$ is small and shows a weak $T$ dependence, 
$d\rho _{ab}/dT$ exhibits a rapid increase below $T\sim $ 10 K close to 
the low temperature scale of Kondo lattice behavior, $T^*\sim$ 8 K. 
This suggests that further coherence develops among $f$ electrons due to the formation of heavy quasi particles below this temperature.  
The absence of similar increase in $d\rho _c/dT$ and large $\rho _{ab}$/$\rho _{c}$ suggest that 
the associated heavy fermions are only
mobile within the $ab$-plane, but not along the $c$-axis. 
This is consistent with the recent band calculation which found that the dispersion along the $ab$-plane is
narrow due to the 4$f$ electron contribution in comparison with the one along the $c$-axis for many of the bands 
mostly coming from conduction electrons~\cite{Tompsett10}. 
We find the anomalies in $d\rho /dT$ at 40-50 K around the same  
temperature scale as for the reflection points in $\chi$(Fig. \ref{f1} (b)) 
where $\chi$ starts to show further increase on cooling. 
This temperature scale can be regarded as the onset temperature of the Kondo lattice behavior.  

A possible explanation for the large anisotropy would be the anisotropic hybridization between the conduction and $f$ electrons
\ie the smaller hybridization along the $c$-axis. 
In this case, while $T_0\sim$ 200 K has its origin 
in the in-plane hybridization, the hybridization scale along the $c$-axis should be smaller.
This may also explain why the coherence peak is observed only in $\rho _{ab}$. 
Indeed, the recent band calculation suggests the smaller hybridization along the $c$-axis in \ybal~\cite{Tompsett10}. 
Although the lower symmetry in \aybal makes its band structure more complex, 
the general features such as anisotropic hybridization are expected to be similar to each other. 

In addition, according to a recent theory on the electronic structure, 
a hybridization node is expected along the $c$-axis in $\alpha$- and \ybal based on the local Yb site symmetry 
if the crystal field ground doublet is made solely of $|J_z = \pm 5/2>$~\cite{Coleman_private, Andriy09}. 
In this case, the $c$-axis transport should mostly come from the conduction electrons and thus 
should have much larger conductivity because of almost no scattering by $f$-electrons. 
The resultant anisotropy of the resistivity should be large when the ground 
$|J_z = \pm 5/2>$ state is dominant at low temperatures. 
If $f$ electrons start populating the excited CEF levels on heating, 
the ratio $\rho _{ab}/\rho _c$ should decrease because the node is no longer well defined. 
Indeed, as shown in Fig. \ref{f4} (a), $\rho _{ab}/\rho _c$ has a large 
value below $\sim$ 10 K, and rapidly decreases with a characteristic temperature scale close to 
the CEF gap energy of $\sim$ 80 K~\cite{Andriy09}. 

The above two theoretical indications strongly support the existence of the anisotropic hybridization. 
It is not likely that the Kondo lattice scale $T^* \sim$ 8 K comes from 
the smaller hybridization scale along the $c$-axis because no feature is observed at $\sim$ 8 K in $\rho _c$. 
Instead, as it is already discussed, $T^* \sim$ 8 K should arise from the in-plane correlation among $f$ electrons. 
To confirm this, Yb-Yb intersite correlation effect should be clarified through, for example, 
the Lu dilution study in Yb$_{1-x}$Lu$_x$AlB$_4$ systems.


\begin{figure}[tb]
\begin{center}
\includegraphics[width = 7.7 cm, clip]{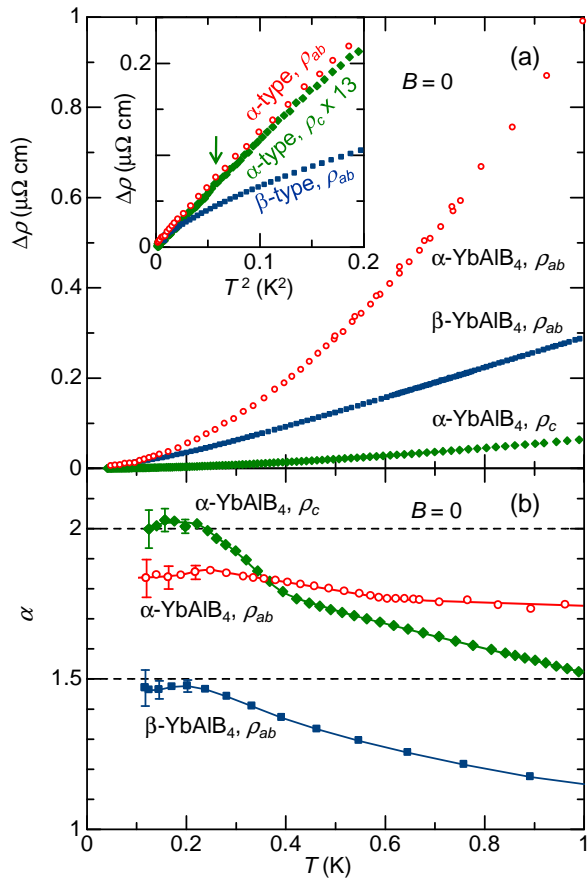}
\end{center}
\caption{(a) Temperature dependence of $\Delta \rho \equiv \rho - \rho _0$.  
Inset shows $\Delta\rho$ versus $T^2$. Note that $\Delta\rho _{c}$ for \aybal is multiplied by a factor of 13 for clarity.
The arrow indicate $T_F$ = 240 mK estimated for $\rho _c$ of \aybal by using the resistivity exponent $\alpha$(see text).
(b) The resistivity exponent $\alpha$ defined by $\Delta \rho = AT^{\alpha}$(see text). }
\label{f5}
\end{figure}

The temperature dependent parts of the resistivity $\Delta\rho \equiv \rho - \rho _0$ at $T <$ 1 K are shown in Fig. \ref{f5} (a). 
Here $\rho _0$ is the zero temperature limit of the resistivity, which was estimated 
by a power law fit of the low $T$ data down to 35 mK(the detail is discussed later). 
$\rho _0$ are 9.4, 0.82 $\mu\Omega$cm for $\rho _{ab}$, $\rho _{c}$ of \aybal, respectively (RRR $\sim$ 20 and 70)
and 0.49 $\mu\Omega$cm for $\rho _{ab}$ of \ybal (RRR $\sim$ 250). 
The anisotropy in $\rho _0$, which corresponds to $\rho _{ab}$/$\rho _{c}\sim 11$ in the lowest $T$, 
is almost the same as that of $\Delta\rho$ ($\Delta\rho _{ab}$/$\Delta\rho _{c}$) which is as large as 13 
in the low $T$ limit (Fig. \ref{f4} (a)).  
$\Delta\rho _{ab}$ of \ybal takes a value between $\Delta\rho _{ab}$ and $\Delta\rho _{c}$ of the $\alpha$-phase. 
On the other hand, if we compare $\Delta\rho$/$\rho _0$ 
, \ybal exhibits much larger value than those in \aybal. 
For instance, $\Delta\rho _{ab}/\rho _0 = 0.85$ at $T =$ 1 K in \ybal is $\sim$ 10 times larger than the respective 
$\Delta\rho /\rho _0$ =  0.11($\rho _{ab}$) and 0.08($\rho _{c}$) for \aybal. 
This cannot be explained only by the better sample quality in \ybal, and thus the  
quantum criticality in \ybal should also be responsible for the enhancement. 
Indeed, the application of the magnetic field, suppressing the criticality, 
decreases $\Delta\rho /\rho _0$ of \ybal to the same order as that in \aybal. 
Note that even a \aybal sample with the highest RRR $\sim$ 110 (estimated by $\rho _c$) does not exhibit superconductivity down to 35 mK (not shown). 

To demonstrate the difference in the ground state of $\alpha$- and \ybal, 
we show the temperature dependence of the power law exponent $\alpha$ defined by $\Delta\rho = \rho _0 + A'T^{\alpha}$ (Fig. \ref{f5} (b)). 
$\alpha$ is obtained by using the equation $\alpha = d\log \Delta\rho / d\log T$. 
$\rho _0$ was determined using the best fitting result to the above equation that indicates the corresponding power law behavior 
in the widest temperature range from the lowest temperature.  
$\alpha$ is strongly dependent on $\rho _0$, and 
its error due to 0.01\% change in $\rho _0$ are shown in Fig. \ref{f5} (b).   
While the exponent $\alpha$ in \ybal is small $\lsim$ 1.5 at the low temperatures, those in \aybal are much larger and 
approaches the normal value of 2 expected for FL on cooling. 
This can be also confirmed in the plot against $T^2$ (inset of Fig. \ref{f5} (a)), where $\rho _{c}(T)$ of \aybal shows 
a linear dependence on $T^2$ below $T_{\rm FL}\sim$240 mK. 
The observation of $\alpha\sim$ 2 in the lowest temperatures in addition to almost saturating $\chi$ and $C_{\rm m}/T$ below $T^*\sim$ 8 K 
indicates that the ground state of \aybal is a Fermi liquid.

\begin{figure}[tb]
\begin{center}
\includegraphics[width = 8.0 cm, clip]{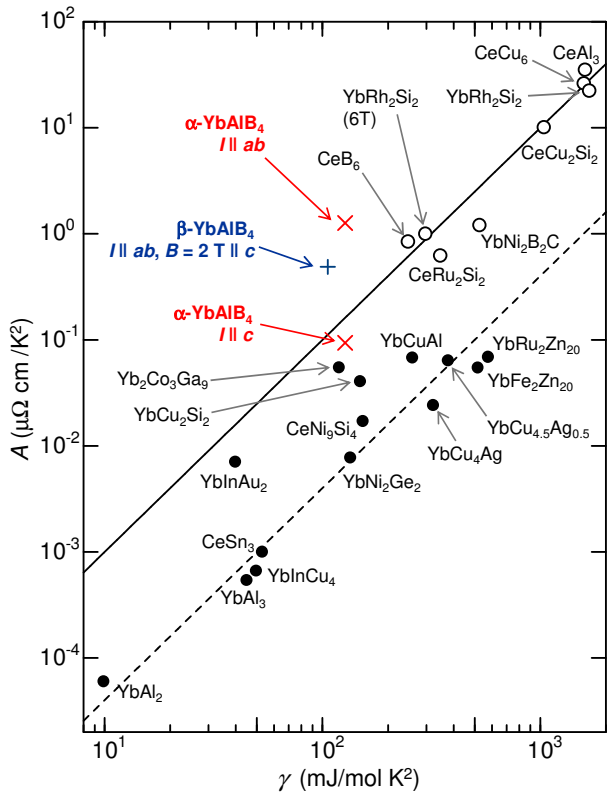}
\end{center}
\caption{$T^2$-coefficient $A$ of the resistivity
versus the $T$-linear coefficient of the specific heat $\gamma$ for both $\alpha$- and \ybal as well as 
for other Ce and Yb based heavy fermions~\cite{tsujiiPRL05, Torikachvili07, matsuda-PSS10}. 
The open circles denote Kondo lattice systems or the systems with crystal field ground state degeneracy $N=2$. 
The closed circles denote 
mixed valent systems, Ce systems with $N=6$ and Yb systems with $N=8$.  
The solid line indicates the original Kadowaki-Woods ratio $A/\gamma ^2=1.0$ $\times 10^{-5}$ $\mu\Omega$ cm(K mol/mJ)$^2$. 
The broken line corresponds to $A/\gamma ^2=4.0$ $\times 10^{-7}$ $\mu\Omega$ cm(K mol/mJ)$^2$, which is the typical value in  
transition metals~\cite{rice68}. 
The data of \ybal at $B=2$ T $\parallel c$~\cite{nakatsuji08, matsumoto-ZFQCP} is also shown.
}
\label{f6}
\end{figure}

The $T^2$-coefficient $A$ defined by $\Delta\rho = \rho _0 + AT^2$ was estimated 
by the linear fit in the inset of Fig. \ref{f5} (a) below 240 mK. The obtained $A$ values are 
0.094 and 1.27 $\mu\Omega$ cm / K$^2$ for $\rho _{c}$ and $\rho _{ab}$, respectively. 
Kadowaki-Woods ratio $A/\gamma ^2$ estimated by using these anisotropic $A$ values are 
5.8 $\times 10^{-6}$ and 7.8 $\times 10^{-5}$ $\mu\Omega$ cm(K mol/mJ)$^2$ for $\rho _{c}$ and $\rho _{ab}$, respectively.  
Here $\gamma$ is a low temperature limit of $C/T$, and in the present case, the value at 0.4 K (127 mJ/mol K$^2$) was used. 
It is known that, the ratio 
$A/\gamma ^2$ is close to 1.0 $\times 10^{-5}$ $\mu\Omega$ cm(K mol/mJ)$^2$ 
in many heavy fermion compounds of Kondo lattice systems~\cite{Kadowaki-Woods}. 
On the other hand, Tsujii \etal  ~have suggested that the ratio is considerably smaller  
in intermediate valence systems, or equivalently, 
the systems with large orbital degeneracy $N$ \ie the system with a large hybridization scale 
$T_0$ compared to CEF splitting $\Delta$~\cite{tsujii03, tsujiiPRL05}. 
In this case, the expected ratio is close to the typical value known for   
transition metals $A/\gamma ^2=0.4$ $\times 10^{-6}$ $\mu\Omega$ cm(K mol/mJ)$^2$, which is 25 times smaller than 
the above ratio for heavy fermions~\cite{rice68}. 
To illustrate this, we show in Fig. \ref{f6}, a full logarithmic plot of $A$ versus $\gamma$ (Kadowaki-Woods plot) for 
representative Ce and Yb based 4$f$ electron systems~\cite{tsujiiPRL05, Torikachvili07, matsuda-PSS10}.
$A/\gamma ^2$ for most of the mixed valent materials or materials with large $N$ (in Ce systems: $N=6$ and in Yb systems: $N=8$) is  
much smaller than the original Kadowaki-Woods ratio and has value of the order of $10^{-7}$ $\mu\Omega$ cm(K mol/mJ)$^2$. 
Compared to these small values observed in mixed valence materials, 
the ratio obtained for $\rho _{c}$ and $\rho _{ab}$ of \aybal is much larger and close to the typical value for heavy fermions.  
In \ybal, the ratio for $\rho _{ab}$ also takes a similar value of 
4.4 $\times 10^{-5}$ $\mu\Omega$ cm(K mol/mJ)$^2$ in magnetic field of 2 T along the $c$-axis~\cite{nakatsuji08, matsumoto-ZFQCP}. 
The large $A/\gamma ^2$ in \aybal and \ybal ($B$ = 2 T $\parallel$ $c$-axis) indicate 
that the system behaves like Kondo lattices at low temperatures rather than 
mixed valent materials.  
Interestingly, the ratio obtained for $\rho _{ab}$ in both $\alpha$- and \ybal is several times larger than the typical value 
for Kondo lattice systems. 
This deviation may come from material dependent properties such as dimensionality and carrier concentration~\cite{jackoNphys5}. 
Further analyses based on fermiology is required to clarify the origin of the enhancement in the Kadowaki-Woods ratio.

\section{Conclusion}
Our detailed measurements have confirmed 
that both \aybal and \ybal 
exhibit Kondo lattice behavior with 
a small renormalized temperature scale of $T^*\sim 8$ K in addition to a large valence fluctuation scale of $\sim 200$ K. 
Below $T^*\sim 8$ K, \aybal forms a heavy Fermi liquid state with $\gamma\sim$ 130 mJ/mol K$^2$ in contrast to the unconventional quantum criticality observed in \ybal.  
The Kadowaki-Woods ratio takes a typical value for Kondo lattice systems and considerably larger than those for  
mixed valent systems. 
This is consistent with the Kondo lattice behavior found in the temperature dependence of the susceptibility and specific heat. 
The large Wilson ratio more than 7 suggests that a ferromagnetic intersite 
coupling between Yb $4f$-electrons and / or proximity to a valence quantum criticality,
may be the origin of the Kondo lattice behavior.  
Furthermore, the large anisotropy observed in the resistivities suggests that hybridization 
between 4$f$ and conduction electrons is much stronger 
within the $ab$-plane than along the $c$-axis. 
This strongly anisotropic hybridization and the large Wilson ratio 
are the keys to understand the unusual Kondo lattice behavior and heavy fermion formation 
in these mixed valence compounds.
The future works including 
neutron scattering measurements and studies of Lu dilution effect in Yb$_{1-x}$Lu$_x$AlB$_4$ systems 
are necessary to clarify the origin of these unusual behaviors.

\begin{acknowledgments}
We thank N. Horie, E. C. T. O'Farrell, C. Petrovic, P. Coleman, A. H. Nevidomskyy, H. Harima, 
S. Watanabe, C. Broholm, K. Ueda 
and T. Sakakibara for supports and 
useful discussions.
This work is partially supported by Grants-in-Aid (No.
21684019) from JSPS, by
Grants-in-Aids for Scientific Research on Innovative Areas (No. 20102007, No. 21102507) from MEXT, Japan, by Global
COE Program ``the Physical Sciences Frontier", MEXT, Japan, by Toray Science
and Technology Grant.
\end{acknowledgments}

%

\end{document}